# Threats and Security Strategies for IoMT Infusion Pumps

Ramazan Yener[1] Muhammad Hassan[2] and Masooda Bashir[3]

[1]University of Illinois Urbana- Champaign
[2]University of Illinois Urbana- Champaign
[3]University of Illinois Urbana- Champaign

**ABSTRACT**

The integration of the Internet of Medical Things (IoMT) into healthcare systems has transformed patient care by enabling real-time monitoring, enhanced diagnostics, and enhanced operational efficiency. However, this increased connectivity has also expanded the attack surface for cybercriminals, raising significant cybersecurity and privacy concerns. This study focuses on the cybersecurity vulnerabilities of IoMT infusion pumps, which are critical devices in modern healthcare. Through a targeted literature review of the past five years, we analyzed seven current studies from a pool of 132 papers to identify security vulnerabilities. Our findings indicate that infusion pumps face vulnerabilities such as device-level flaws, authentication and access control issues, network and communication weaknesses, data security and privacy risks, and operational or organizational challenges that can expose them to lateral attacks within healthcare networks. Our analysis synthesizes findings from seven recent studies to clarify how and why infusion pumps remain vulnerable in each of these areas. By categorizing the security gaps, we highlight critical risk patterns and their implications. This work underscores the scope of the issue and provides a structured understanding that is valuable for healthcare IT professionals and device manufacturers. Ultimately, the findings can inform the development of targeted, proactive security strategies to better safeguard infusion pumps and protect patient well-being.

## INTRODUCTION

The Internet of Medical Things (IoMT), a specialized subset of the Internet of Things (IoT), is revolutionizing the healthcare sector by enabling the seamless connection, interaction, and data exchange between medical devices and healthcare networks. According to the National Institute of Standards and Technology (NIST), IoT encompasses the network of devices that contain the hardware, software, firmware, and actuators which allow the devices to connect, interact, and freely exchange data and information." Within healthcare, IoMT devices include medical imaging systems, infusion pumps, wearable biosensors, and other connected tools that facilitate remote monitoring and enhance clinical workflows (Das 2022a).

IoMT is transforming healthcare delivery by reducing costs, improving efficiency, and enabling remote patient monitoring. These advancements are particularly valuable for chronic disease management and rural healthcare access. Devices such as infusion pumps have become integral to daily clinical practices due to their ability to deliver fluids and medications with precision (O'Brien et al. 2018). However, the increasing reliance on IoMT also introduces significant cybersecurity risks. Infusion pumps, for instance, are vulnerable to a range of threats such as unauthorized access, data breaches, and malicious cyber-attacks that could jeopardize patient safety and interfere with healthcare operations.

Securing IoMT infusion pumps is essential for ensuring patient safety and preserving the reliability of healthcare systems. A compromised infusion pump can lead to life-threatening consequences by altering medication dosages or disrupting critical care workflows. Additionally, breaches involving Protected Health Information (PHI) expose healthcare providers to legal liabilities and reputational damage. Addressing these risks requires a comprehensive understanding of the vulnerabilities inherent in these devices and effective mitigation strategies tailored to their unique operational contexts.

## BACKGROUND

This work aims to provide context for the rapidly growing field of the Internet of Medical Things (IoMT) in healthcare, with a particular focus on infusion pumps. We summarize prior research regarding IoMT applications, highlight common security gaps in these devices, and discuss existing studies on IoMT security, emphasizing the research gap addressed in this paper.

### IoMT Overview

IoMT market is rapidly expanding with estimates forecast that it will reach approximately $170 billion by 2030 (Grand View Research 2023). Moreover, forecasts suggest that by 2026, smart hospitals around the globe will implement 7.4 million connected IoMT devices averaging more than 3,850 devices per hospital. Press (2022)- with additional personal IoMT devices, such as smart wearables, contributing to further growth.

Applications of IoMT span a range of healthcare functions. For example, these devices facilitate continuous monitoring of patients' vital signs, enabling real-time data collection and prompt identification of potential health risks. In elderly care, technologies such as smart belts utilize airbags to mitigate fall-related injuries (Tarbert & Singhatat 2023). Similarly, Empatica's smartwatch is used to detect early signs of seizures in individuals with epilepsy, alerting both patients and caregivers to enable timely intervention. Studies have demonstrated that such devices can achieve clinically acceptable levels of accuracy (Gerboni et al. 2023). Moreover, seamless integration of IoMT data into electronic health records (EHRs) boosts data accuracy and accessibility, leading to improved patient care and more informed clinical decision-making (Lutkevich n.d.).

Smart infusion pumps represent a critical component of IoMT in healthcare. These sophisticated devices incorporate features such as Dose Error Reduction Systems (DERS) and customizable drug libraries to improve the precision of medication delivery. Although infusion pumps are widely adopted in clinical settings, challenges remain in optimizing their use through workflow integration, process improvements, and continued training (Hoffman & Bacon 2020). Additionally, detailed descriptions of related devices for instance, insulin pumps reveal a blend of mechanical and electronic components that harmonize to ensure precise drug dosing (Singh et al. 2021).

Previous studies highlight risks from unsecured medical devices, such as unauthorized data access and interference with critical interventions (U.S. Department of Health & Human Services - Office for Civil Rights n.d., Proofpoint

32024). Broader investigations show that the growing volume of data and interconnected systems in IoMT heighten threats like network overload and data breaches (Osama et al. 2023).

Ahmed et al. (2023) emphasize the need to improve data collection, protection, and storage to detect and mitigate threats. Despite technological advances, vulnerabilities persist including insecure electrocardiographs, compromised imaging systems, and unprotected dispensers (Vedere Labs 2024). While these works address IoMT security generally, comprehensive categorization and targeted mitigation strategies for infusion pumps remain limited. Das et al. (2023) point the urgency of building robust security frameworks tailored to IoMT.

### Known Vulnerabilities

Despite the numerous benefits, IoMT devices including smart infusion pumps face substantial cybersecurity challenges. Common vulnerabilities include weak encryption, poor authentication protocols, and inadequate network segmentation (Dzamesi & Elsayed 2025). Many devices operate with outdated software or firmware and default credentials, leaving them exposed to unauthorized access and potential sabotage. One study examining connected devices identified over 4,000 vulnerabilities; approximately 2% of these were found in IoMT devices, with 80% considered critical (Labs 2023).

Specifically for infusion pumps, known security gaps involve unpatched software, insecure communication protocols, and the potential for physical access exploits (Davis 2021, Nair 2022). An analysis of more than 200,000 infusion pumps indicated that 75% had identifiable cybersecurity weaknesses, with over half vulnerable to high-severity exploits as early as 2019 Das (2022b). Additionally, proper security measures such as robust authentication, encryption, and regular firmware updates are often lacking, making these devices prime targets for cyber-attacks (Guarding the flow: Enhancing cybersecurity measures for infusion pump systems 2024).

Though infusion pumps offer clear clinical benefits, their cybersecurity challenges demand continued research. This paper contributes a focused analysis of infusion pump vulnerabilities and outlines security measures to secure IoMT infrastructure.

## METHODOLOGY

### Literature Search and Selection

A comprehensive literature review was conducted to identify cybersecurity vulnerabilities specific to infusion pumps. We searched five academic databases: Scopus, Web of Science, Google Scholar, IEEE Xplore, and ACM Digital Library.

### Search Strategy

The search used the keywords "infusion pumps" and "vulnerability." For IEEE Xplore and ACM Digital Library, we applied "infusion pumps" and "security vulnerability," limiting results to 2020–2025. In Google Scholar: "infusion pumps" AND "security vulnerability" AND "IoMT" ("smart infusion pump" OR "drug infusion system") AND ("cyber threat" OR "hacking" OR "attack surface") AND ("healthcare" OR "medical device"), also within the 2020–2025 range. ACM Digital Library was searched using: [[Abstract: "infusion pump"] OR [Abstract: "insulin pump"]] AND [Abstract: vulnerability]. IEEE Xplore was queried with:

("All Metadata": infusion pump) AND ("All Metadata": vulnerabl*) AND ("All Metadata": cybersecurity). Scopus and Web of Science were both searched using: ("infusion pumps" AND "vulnerability").

### Inclusion and Exclusion Criteria

In Scopus, terms were applied to the "Article Title, Abstract, and Keywords" fields; in Web of Science, within the "Topic" field. All databases were filtered for publications from 2020–2025.

Included studies were peer-reviewed articles and conference papers explicitly addressing cybersecurity vulnerabilities in infusion pumps. Excluded materials included: Dissertations, Irrelevant papers (e.g., clinical studies without a security focus), Documents without full-text access, Duplicate entries, Grant announcements or abstracts lacking technical detail, Review papers.

### Final Selection

From an initial pool of 132 papers, seven met the selection criteria after rigorous screening. Table 1 summarizes the search results from each database

| Database | Search Results | Relevance | Selected Papers |
|---|---|---|---|
| Scopus | 28 | 5 | 1 |
| Web of Science | 18 | 4 | 1 |
| Google Scholar | 77 | 17 | 3 |
| IEEE Xplore | 6 | 3 | 1 |
| ACM Digital Library | 3 | 1 | 1 |
| Total | 132 | 30 | 7 |

**Table 1.** Search Results from Various Databases

### Data Extraction and Analysis

The analysis focused on identifying the most commonly reported vulnerabilities. The following five security issues emerged as the primary concerns discussed in the literature: device-level flaws, authentication weaknesses, network vulnerabilities, data privacy risks, and operational challenges.

## FINDINGS

This section describes the main outcomes of our review, which found five key cybersecurity vulnerabilities in IoMT infusion pumps, summarized below and in table 2.

### Device-Level Vulnerabilities

A prominent issue is buffer overflow, where malicious packet injection bypasses memory protections, enabling arbitrary code execution. Beyrouti et al. (2024) demonstrated this via simulated attacks on medical IoT gateways, causing system crashes and data corruption. Similarly, sensor manipulation attacks, as shown by Chang et al. (2023), exploit adversarial inputs to alter glucose readings in artificial pancreas systems, leading to incorrect insulin dosing. Stergiopoulos et al. (2023) further validated these risks by illustrating how manipulated sensor data disrupts PID controller outputs in insulin pumps. Physical tampering and legacy systems exacerbate these risks. Taylor et al. (2022) documented physical exploits, such as replacing WiFi cards with malicious CompactFlash components (CVE-2016-8375), which bypass digital safeguards entirely. Such findings underscore the urgent need for hardware integrity checks and firmware updates to mitigate life-threatening risks.

### Authentication and Access Control Issues

Authentication and access control vulnerabilities in infusion pumps, as reported in the reviewed studies, expose critical risks through weak credential management and insufficient privilege enforcement. These flaws enable unauthorized access to device settings or sensitive patient data, often stemming from design oversights in device firmware and communication protocols.

A recurring issue is the use of hard-coded credentials, which manufacturers embed into device firmware for administrative access. (Taylor et al. 2022) identified static passcodes in Alaris infusion pumps that granted unrestricted access to network configurations, a flaw exploited in real-world scenarios to manipulate device parameters. Similarly, Beyrouti et al. (2024) documented cases where missing authentication checks allowed attackers to bypass critical functions, such as firmware updates, through unvalidated input channels.

Denial-of-Service (DoS) attacks further compound these risks by disrupting sensor functionality. Chang et al. (2023) demonstrated adversarial attacks on Continuous Glucose Monitoring (CGM) sensors, where maxed-out readings caused temporary system failures. Griffy-Brown et al. (2022) corroborated this by highlighting code injection vectors that exploit weak authentication to overload device resources. Such attacks not only impair device availability but also create entry points for lateral movement within healthcare networks.

### Network and Communication Vulnerabilities

Network and communication vulnerabilities in infusion pumps arise from insecure protocols and misconfigurations that expose devices to data interception or remote exploitation. These weaknesses, reported in the reviewed literature, highlight systemic risks in healthcare networks due to the interconnected nature of IoMT devices. For example, a critical issue is clear text data transmission, where sensitive information, such as sensitive data configurations or patient data, is exchanged without encryption. Beyrouti et al. (2024) demonstrated this vulnerability through simulated attacks on medical IoT gateways, showing how unencrypted communication channels allow adversaries to intercept and manipulate data via man-in-the-middle (MITM) attacks. Likewise, Mejía-Granda et al. (2024) identified FTP server exploitation in syringe pumps, where unsecured file transfer protocols enabled unauthorized access to device settings or firmware. The consequences of such vulnerabilities extend beyond individual devices. For instance, compromised infusion pumps can serve as entry points for lateral movement within hospital networks, enabling attackers to breach Electronic Health Records (EHRs) or disrupt clinical workflows. Taylor et al. (2022) further emphasized risks like credential theft via physical network card replacement (CVE-2016-8375), illustrating how hardware-level flaws compound network insecurities.

### Data Security and Privacy Risks

Data security and privacy risks in infusion pumps involve exposure or unauthorized access to protected health information (PHI), as reported across multiple studies. Beyrouti et al. (2024) demonstrated how unencrypted communication between infusion pumps and IoT gateways allows adversaries to intercept sensitive data via man-in-the-middle (MITM) attacks. Similarly, Mejía-Granda et al. (2024) identified sensitive information leakage in syringe pumps, where unsecured FTP protocols exposed PHI, including patient identifiers and treatment details. Taylor et al. (2022) documented hard coded passwords in Alaris pumps that granted unrestricted access to network settings, allowing attackers to exfiltrate PHI or manipulate device parameters and show that insufficiently protected credentials further exacerbate these risks. These vulnerabilities, stemming from insecure data handling practices, threaten patient confidentiality and institutional compliance with healthcare regulations.

### Operational and Organizational Challenges

Operational and organizational challenges in securing infusion pumps stem from systemic issues such as outdated software, misconfigurations, and inadequate maintenance practices, as reported in the reviewed studies. Mejía-Granda et al. (2024) identified outdated operating systems (e.g., Windows XP) in medical devices, which lack modern security features and leave pumps exposed to unpatched vulnerabilities. The study also highlighted security misconfigurations, such as improper network segmentation and disabled encryption policies, which enable attackers to bypass safeguards. For instance, Taylor et al. (2022) documented maintenance mode vulnerabilities in Alaris pumps, where unprotected diagnostic interfaces allowed unauthorized access to error logs and network configurations. These challenges create persistent vulnerabilities that adversaries can exploit over extended periods.

## DISCUSSION

Despite growing awareness, many of the same fundamental weaknesses persist across studies and device models. A consistent pattern emerges: vulnerabilities span multiple layersfrom insecure firmware and default passwords to exposed networks and weak governance. Poor authentication practices stand out in nearly every source. Both targeted case studies by Taylor et al. (2022) and broad surveys by Mejia-Granda et al. (2024) report hard-coded or default credentials, signalling a systemic failure to address even basic access controls. Similarly, the frequent use of outdated software and unpatched flaws points to an industry lag in keeping medical devices secure against evolving threats. When one study finds 75% of infusion pumps have known vulnerabilities and another confirms active CVEs on widely used models, it's clear that these are not isolated issues but part of a broader, ongoing challenge.

Another recurring problem is insecure communication and integration. Several studies identify unencrypted channels, open ports, and flat network configurations as major risks as shown by Taylor et al. (2022) & Mejia- Granda et al. (2024). These network gaps often connect directly to other vulnerabilities: weak credentials can be exploited remotely due to open access, leading to data exposure and long-term compromise. The reviewed literature spanning technical tests and policy analysis shows how these issues reinforce one another. Even advanced devices with AI features, such as those discussed by Chang et al. (2023) and Stergiopoulos et al. (2023), remain vulnerable if core weaknesses like access control are not addressed Chang et al. (2023) Stergiopoulos et al. (2023).

Overall, infusion pump cybersecurity demands a coordinated approach. The studies reviewed recommend building security into device design, enforcing strong authentication and encrypted communication, segmenting networks, and promoting proactive risk management in clinical settings Taylor et al. (2022)Mejía-Granda et al. (2024) Beyrouti et al. (2024). By categorizing vulnerabilities across technical and organizational dimensions, this review connects isolated findings into a more complete picture. A weakness identified in one context often echoes broader trends across the field. By drawing out these connections, this synthesis helps inform more holistic strategies for improving the security of these critical healthcare devices.

## Limitations

One of the limitations of our study is relying on a small number of recent papers. Although these papers are insightful it may not encompass the full range of perspectives in infusion pump vulnerabilities. Key perspectives from industry technical reports or a wider variety of device types is not included. As a result, our brief review and synthesized findings does skew towards the particular vulnerabilities and solutions emphasized in these selected works. Additionally, We focused primarily on infusion pumps and closely related hospital devices. This narrower focus, while keeping the analysis on track, limits the breadth of our implications.

## CONCLUSION

Our review shows that IoMT infusion pumps face a broad set of cybersecurity vulnerabilities across five main areas: device-level flaws, authentication and access control weaknesses, network and communication issues, data security risks, and operational shortcomings. These vulnerabilities often intersect poor access control can open paths for network intrusions that compromise device behaviour or leak sensitive information. The consequences are serious: a compromised pump could alter medication delivery or serve as a foothold for broader attacks within healthcare systems. To reduce these risks, organizations need to implement layered defences, such as removing hard-coded credentials, maintaining up-to-date software, securing communication channels, and isolating devices from general networks. These technical efforts must be backed by institutional practices like staff training and integration of device security into broader risk management. The findings across the seven studies underscore the need for both technical and organizational action. As infusion pumps remain vital to patient care, securing them should be considered a core safety concern. Continued collaboration among manufacturers, healthcare stakeholders, and regulators will be key to strengthening standards, improving designs, and ensuring these devices are resilient in today's connected healthcare environments.